\documentstyle[preprint,aps]{revtex}
\begin{document}
\title{Nuclear Magnetic Relaxation in Sc Metal}
\author{M.Hondoh, I.R.Mukhamedshin, M.Nasu, H.Suzuki}
\address{Faculty of Science, Kanazawa University, Kanazawa, 920-11, Japan}
\author{A.V.Klochkov, V.V.Naletov, R.N.Romanov, M.S.Tagirov, D.A.Tayurskii}
\address{Physics Department, Kazan State University, Kazan, 420008, Russia}
\date{\today}
\maketitle

\begin{abstract}
The nuclear magnetic relaxation in scandium metal is investigated at liquid
helium temeprature for both single crystal and powder samples. The obtained temperature
and field dependence of the Korringa constant are described by the means
of the spin fluctuation theory. It is shown that one of the reason for the
high field decreasing of the relaxation rate is the quenching of spin fluctuations.
The other possible reason is the magnetic impurity influence.
The reasons of the existing variations of the Korringa constant are discussed. The possible sources
of non-single exponential relaxation are investigated.
\end{abstract}

\section{\protect\bigskip Introduction}

The interest for investigations of the behavior of nuclear spin system in
scandium $^{45}Sc$ ($I=7/2$) metal comes from two sides at least. First of
all, scandium is the first of transition metal and has only one $3d$
electron. But among transition metals possessing unique magnetic properties
as a result of their $3d$ electrons scandium is probably the least
understood, although in principle with only one $d$ electron it might be
expected to be one of the best understood. The reason for such an absence of
information is the impossibility before recent years \cite{Gschneidner} to
get high purity scandium metal, especially with respect to the iron content.

Secondly, scandium is one of the metals where it could be possible to get
and to investigate the nuclear magnetic ordering \cite{Suzuki1},\cite
{Suzuki2}. Due to the smallness of the Korringa constant the nuclear
spin-relaxation time is so short that the temperature of the conduction
electrons and the temperature of nuclear spins are practically equal during
experiment. It means that the exchange interaction between nuclear spins is
rather strong that should lead to rather high temperatures of nuclear
ordering (order of $\mu K$ while in such metals as $Cu$, $Ag$ and $Rh$ the
corresponding temperatures have the order of few $nK$ and even of hundreds $%
pK$ \cite{RMP}). Nuclear magnetic resonance (NMR) besides the knowledge of the
parameters of nuclear spin system yields information about conduction
electron subsystem not immediately deducible from the band structures. In
particular, the core electrons and the valence electrons give the distinct
contributions to the Knight shift and nuclear magnetic relaxation time.

Accordingly to the above-mentioned it becomes to be clear the validity of
the correct determination of the nuclear spin-lattice relaxation time from
measurements of nuclear longitudinal magnetization evolution. Not so
extensive NMR investigations in $Sc$ metal \cite{Blumberg}-\cite{Masuda3}
in the temperature range $0.001\div 300K$ gave  rather different values for
the Korringa constant $K$ (from $0.09\sec \cdot K$ to $1.6\sec \cdot K$) in
the dependence on experimental conditions, particularly on the applied
magnetic fields (see Table 1). It should be noted that in zero magnetic
field (nuclear quadrupole resonance investigations \cite{Cornell}) the value
of the Korringa constant obtained from the longest magnetization decay is
extremely low, namely $0.09sec\cdot K$. A priori it might be expected that
such magnetic field dependence of the Korringa constant could be explained
as the result of spin fluctuation influence. It is well known that the heat
capacity at high magnetic fields is decreasing due to the spin fluctuation
quenching \cite{SF}. As for nuclear magnetic relaxation such influence
was observed for $Sc$ intermetallic alloys \cite{Sc3In}. But in the case of $%
Sc$ pure metal there is no information about spin fluctuation influence on
the nuclear spin system. It should be noted here that the recent theoretical
calculations of spin relaxation times and susceptibilities on the basis of
magnetic correlation functions \cite{Gotz} showed the importance of taking
into account the properties of Fermi surface as well as the band structure
in a finite energy range. Nevertheless in the framwork of this approach it
is impossible to explain the variation of the Korringa constant in the
different experiments.

The present work reports about long time experimental and theoretical
investigations of nuclear magnetic resonance in pure scandium metal (powders
and single crystal). The special attention has been payed to the possible
sources for non- exponential relaxation such as non-complete saturation of
NMR line, the heating of conduction electron subsystem and influence of spin
fluctuations.

\section{Samples.}

The purity of our powder sample which we bought from Rare Metallic Co., Ltd. 
is 3N. 
A single crystal used in our experiments was grown in Ames Laboratory. 
It has a dimension of 3.5x2.0x27 $mm^3$ with the long direction parallel to the a-axis. 
The main magnetic impurities in our specimen analyzed in Ames Laboratory are $3 at. ppm$ $Fe$, 
$0.23 at. ppm$ $Cr$ and $3.2 at. ppm$ $Mn$.

\section{NMR spectra.}

Because of hexagonal symmetry NMR spectrum of $^{45}Sc$ ($I=7/2$ ) should
consist of the main central peak with six sattelites. The anisotropy of the
Knight shift \cite{Barnes} leads to asymmetrical form of each sattelite line
as well as central one. In Fig.\ref{fig1} the NMR spectrum of $Sc$ polycrystalline
sample at liquid helium temperature ($T=1.5K$) is shown. We were not able to
observe all satelite lines. Namely the lines corresponding to the
transitions $\frac{7}{2}\leftrightarrow \frac{5}{2}$ and $-\frac{5}{2}%
\leftrightarrow -\frac{7}{2}$ were not detected. It is connected with the
sensivity of our apparatus as well as with the polycrystalline sample
preparation. Only very careful preparation of metal filings allowed to
obesrve all six satellite lines in a powder sample\cite{Barnes}. It was shown in \cite
{Jones} that the spacings between corresponding satellite pairs in 
NMR linshape in metals are
unaffected by the anisotropic Knight shift through the second order of
perturbation theory, and are also independent of the second-order quadrupole
contributions: 
\begin{equation}
\nu \left( -m+1\longleftrightarrow -m\right) -\nu \left(
m\longleftrightarrow m-1\right) =\left( m-\frac{1}{2}\right) \nu _{Q},
\label{Q}
\end{equation}
where $\nu _{Q}=3e^{2}qQ/2I(2I-1)h$ is a convinient measure of the
quadrupole interaction strength. Here $e$ is the charge of electron, $eQ$ is
the electric quadrupole moment of the nucleus, $eq$ is the total electric
field gradient at the nucleus, $I$ represents the angular momentum quantum
number and $h$ is Planck's constant. We obtained from our data $\nu _{Q}\approx
0.128MHz$ that agrees well with the value $v_{Q}=0.144MHz$ reported in \cite
{Barnes} for polycrystalline sample, with the value $nu _{Q}=0.124MHz$
obtained in \cite{Ross} from nuclear resonance measurements at single
crystal sample and with the value $v_{Q}=0.13MHz$ measured in \cite{Cornell}
by nuclear quadrupole resonance method.

The {\it cw} NMR of scandium is the first important step if one is going to
provide the correct measurements of spin-lattice relaxation by using pulse
NMR technique. Only in this case one can estimate the pulse duration for
complete saturation of all transitions between nuclear energy levels. It is
well known that NMR spectrum can be obtained by the Fourier transform of
free induction decay (FID) . In Fig.\ref{fig2} one can see FID signals of $^{45}Sc$
in single crystal at different frequences ($6.3$, $4.71$ and $3.65MHz$).
Obviously the scandium FID signal is complicated due to the interference
with copper (containing in NMR coil wires) FID signal. The gyromagnetic
ratio for copper is  rather close to the corresponding value for scandium (
$\frac{\gamma }{2\pi }=1.0348kHz/Oe$ for $^{45}Sc$,  
$\frac{\gamma }{2\pi }=1.128kHz/Oe$ for $^{63}Cu$ and $\frac{\gamma }{2\pi }%
=1.208kHz/Oe$ for $^{65}Cu$), so the lowering of the external magnetic field
leads to the overlapping of all signals.

\section{Relaxation.}

Successfull observation of NMR in metallic single crystals is  rather
seldom because of skin effect. For example, the measurements of NMR in metallic $Tl$ 
\cite{EskaTl}, $Ga$ \cite{EskaGa}, $Al$ \cite{PobellAl}, $Sc$ \cite{Ross} 
samples can be mentioned. By using home made
pulse NMR technique we have observed the spin echo in polycrystalline
samples and single crystal in inhomogeneous applied magnetic field. The
corresponding transverse magnetization decays are shown in Fig.\ref{fig3}. It is
seen that both decays have essential one-exponential character and the
corresponding lineshapes have the Lorentz form. The difference in $T_2$
values can be attributed to the Knight shift dispersion in polycrystalline
samples which leads to more longer value of transverse relaxation time. As
mentioned above that for correct $T_1$ measurements it is very important to
choose the pulse duration for the complete saturation of nuclear spin
system. The Fig.\ref{fig4} illustrates how longitudinal relaxation time 
$T_1$ depends on pulse duration time.
When we use long pulse duration time (the corresponding spectrum is more
narrow) we saturate only some part of nuclear magnetic resonance spectrum.
Consequently the measured $T_1$ is shorter because of spectral diffusion
processes and the relaxation has multi-exponential character (see \cite
{NarathTi} and Section 5). Naturally all our measurements have been provided
by using the shortest pulse duration time. As in case of $T_2$ measurements
the longitudinal magnetization decays have also one- exponential character.
One see it in Fig.\ref{fig5} where longitudinal magnetization evolution were
measured at three different frequencies ($3.65MHz$, $4.71MHz$ and $6.3MHz$).
The obtained values of $T_1$ in our experiments and the data known before 
\cite{Blumberg}-\cite{Cornell} allowed us to plot the magnetic field
dependence of the Korringa constant which is shown in Fig.\ref{fig6}. 

The obtained
magnetic field dependence can be explained by two ways at least. They are
the qeunching of spin fluctuations by magnetic field and the impurity
influence.

As far as
the relaxation rate increasing with magnetic field decreasing is observed,
first of all we can think about magnetic field influence on the spin
fluctuations and work in the framework of
self-consistent renormalization (SCR) theory \cite{MoriyaBook}. Namely it
is known from the heat capacity measurements
that spin fluctuation is quenched at magnetic field above 5 T \cite{SF}.The spin-
lattice relaxation rate of $^{45}Sc$ in intermetallic alloy $Sc_3In$ increases
to almost five times when magnetic field goes down from $6.8T$ to $0.8T$ at
low temperatures \cite{Sc3In}. 

If we restrict ourselves by consideration only Fermi contact interaction
between nuclear and electronic spins then the rate of the nuclear
spin-lattice relaxation is determined by the following expression \cite
{Moriya}: 
\begin{equation}
T_1^{-1}=\gamma ^2A_{hf}^2T\frac{\sum_qIm\left\{ \chi ^{-+}(q,\omega
_0)\right\} }{N^2\omega _0},  \label{T1}
\end{equation}
where $A_{hf}$ describes the hyperfine interaction, 
\[
\chi ^{-+}(q,\omega )=i\int_0^\infty dt\cdot e^{i\omega t}\left\langle
\left[ S^{-}(q,t),S^{+}(-q,0)\right] \right\rangle 
\]
is the transverse electronic dynamic susceptibility. It was assumed here
that an external magnetic field as well as internal one are directed along $%
z $-axis. The calculations of $\chi ^{-+}(q,\omega )$ in the framework of
SCR theory \cite{MoriyaUeda} lead to the formula for the Korringa constant 
\cite{ZrZn2}: 
\begin{equation}
\left( T_1T\right) ^{-1}=A+\frac{B\cdot M/H}{1+C\cdot M^3/H},  \label{SCR}
\end{equation}
which can be re-writen in the form more convenient for the comparison with
the experimental data: 
\begin{equation}
\left( T_1T\right) ^{-1}=A+\frac{B\cdot \varsigma }{1+C\cdot \varsigma
^3\cdot H^2},\text{ }  \label{Korringa}
\end{equation}
where $\varsigma =M/H.$ The values of parameters $A,B$ and $C$ in equation 
\ref{Korringa} can be obtained from the behavior of the Korringa constant at
different magnetic field. So, at high magnetic fields (see for example the
experimental results in \cite{Masuda3}) 
when one approaches to the magnetizaiton saturation limit
the contribution coming from the second 
term in equation 3 will be negligible and we get the value of parameter $A$. 

We should note here that equations 3 and 4 can not be applied for very low
magnetic fields. The well-known Korringa relation hes been obtained in the appoximation
when the interaction between nuclear spins is so small in comparison with the nuclear
Zeeman energy. So it was possible to consider the energy levels for a given nuclear
and to make corresponding calculations. At small magnetic fields one must consider
the system of all nuclear spins and use some thermodynamic relations,
for example, spin temperature concept. Such situation has been realized, for
example, in the measurements of nuclear magnetic relaxation in $Cu$ \cite{Huiku}.
It was shown in \cite{Huiku} that at
low magnetic fields (lower than local magnetic fields) 
magnetic impurity, such as $Fe$, $Cr$, $Ni$, $Mn$, can strongly influence
the nuclear spin-lattice relaxation process.
 
It is important also that at small magnetic fields the effect of electron spin
fluctuations can be partially washed out by effect of non-equal spacings between
nuclear energy levels (for nuclei with quadrupole moments) and by nuclear spin-spin
interactions.

It follows from above that as far as the impurity contents in Cornell' group
experiments \cite{Cornell} was appoximately the same as in our one and these exmperiments were
provided at zero magnetic field (nuclear quadrupole resonance measurements)
we must exclude the Cornell data from our analysis and from the Fig.6.

Generally speaking for the fitting procedure it is necessary
to have the magnetic field dependence of $\varsigma $ parameter which is
magnetization divided by value of the applied magnetic field. The special
experiments at liquid helium temperature with use the SQUID give us the
necessary data shown in Fig.\ref{fig7}. The magnetic field dependence of $\zeta $
parameter is not so strong and can be described by exponential function: 
\begin{equation}
\zeta =a_1+b_1\cdot \exp \left( -\frac{c_1}H\right)  \label{exp}
\end{equation}
as well as by function similar to the pointed out one in equation \ref{Korringa}: 
\begin{equation}
\zeta =a_2+\frac{b_2}{1+c_2H^2}.  \label{quad}
\end{equation}
The results of these fittings are presented in Fig.\ref{fig7} and in Table 2.

Any from the functions \ref{exp} and \ref{quad} can be used in the fitting
procedure for the equation \ref{Korringa}. The result of this fitting is shown
by solid line in Fig. \ref{fig6}. Of course, the agreement between experimental
points and the theory quantitatevly is not so good because of  rather
big differences between the experimental points obtained
by different group. But the tendency to the magnetic relaxation rate increasing
at low magnetic field can be attributed in part to the spin fluctuation quenching.

The another possible reason for the observed increasing is the
impurity influence (see the impurity contents in Section 2).
The magnetic field dependence of spin-lattice relaxation
time can arise from magnetic impurities influence. 
Scandium like palladium
is an exchange enchanced metal and its magnetic properties are very sensitive
to $3d$ impurities. The giant magnetic moments associated with $Fe$ impurities in
a $Pd$ matrix have been observed and their ground state has been investigated in
\cite{Pd}. In the case of scandium there is also possibility for the enhancement
of impuriry magnetic moments by exchange interactions.
The magnetic impurity effect on the magnetic property of $Sc$ metal
were discussed in \cite{Suzuki3} where the observed anomalies below $1mK$ in
magnetic susceptibility and magnetization have been explained as the spin glass
phenomena. It is well known that the magnetic impurities contribution to the nuclear spin-lattice
relaxation rate is proportional to $(1-p^2_0)$, where $p_0$ represents the
equilibrium magnetic polarization of impurity and for two-level system
is equal to $\tanh (\frac{mH}{2k_BT})$. Here $k_B$ is the Boltzman constant, $T$
is temperature and $m$ stands for the magnetic moment of impurity atom. The dashed line in Fig.6 shows
the results of fitting when one takes into account two contributions to the
magnetic relaxation rate  - the magnetic field independent contribution
coming from Korringa mechanism and the contribution from the magnetic
impurities. The value of the impurity magnetic moment obtained in this fitting
procedure is equal approximately to $8 \mu _B$ that two times bigger than
estimations of effective magnetic moment of iron impurity in scandium metal \cite{Suzuki3}.
From this point of view we could conclude that one has to take into account
the all possible mechanisms for magnetic field dependence of nuclear spin-lattice
relaxation in scandium metal. But we should mention here
that the proportionality of the relaxation rate to $(1-p^2_0)$ means that 
Korringa constant also should be temperature dependent. Accordinly to SCR theory
at temperature well above transition temperature into ordered state in electron
system the nuclear relaxation rate $T_1^{-1}$ is approximately proportional to $T\chi$.
The susceptibility of $Sc$ changes a small value with temperature. So
it seems reasonably to think that most dominant field dependence of the Korringa
constant comes from spin fluctuations. Additional measurements at lower temperatures
help to eliminate the influence of spin fluctuations.

\section{The possible sources of non-exponential relaxation}

As it was mentioned above the $^{45}Sc$ NMR spectrum consists of seven
unequally spaced lines: the central line at $\omega _{0}$ corresponding to
the transition $+\frac{1}{2}\longleftrightarrow -\frac{1}{2}$ and six
satellite lines $m\longleftrightarrow m-1$. 
This spectrum is rather wide. For example, at
magnetic field $1.483T$ (the corresponding frequency is $%
15.33MHz$) the
frequencies for the most right and most left satellite lines are
distinguished on the value from $0.5MHz$ to $0.8MHz$ in the dependence on
magnetic field orientaion with respect to crystallographic axis $c$. This
enormous width of the nuclear resonance leads to some difficulties in the
usual pulse NMR experiments in which the recovery of the nuclear
magnetization is observed after application of a saturating ''comb'' of
radiofrequency pulses. It is very difficult to achieve the complete
saturation of spin-system (i.e. to equalize the populations of all nuclear
spin levels) due to a big discrepancies in the resonant frequencies for the
transitions $m\longleftrightarrow m-1$ (for the system of unequally spaced
levels the simultaneous nuclear spin-flips involving two different
transitions are forbidden by energy conservation requirements \cite
{McLaughlin}). The question about the influence of the incomplete saturation
of spin system on the nuclear magnetization recovery in the case of two
isotopes of titanium ($^{47}Ti$, $I=\frac{5}{2}$ and $^{49}Ti$, $I=\frac{7}{2%
}$) has been investigated in \cite{NarathTi}. Here we report only about the
results of our numerical calculations. In Fig.\ref{fig8} the curves of
magnetization recovery are shown at different initial conditions. We assume
here for the numerical calculation purposes the nuclear spin-lattice relaxation time
being equal to $600$ $msec$.
 The curve
1 in Fig.\ref{fig8} corresponds to the case when only central transition $\frac{1}{%
2}\longleftrightarrow -\frac{1}{2}$ is saturated while the other levels have
the equilibrium populations. The curve 2 is obtained for the case when a
comb of sufficient duration is used in order to allow the populations of the 
$\left| m\right| >\frac{1}{2}$ levels to attain the thermal equilibrium with
those of the $\left| m\right| =\frac{1}{2}$ levels. In other words, just
after the comb the $\frac{1}{2}\longleftrightarrow -\frac{1}{2}$ transition
is saturated, while the population differences between all other pairs of
adjacent levels are determined by the lattice temperature. Finally the curve
3 represents the single exponential relaxation observed when all transitions
are saturated at initial time.

But as we will show now the incomplete saturation of NMR line because of its
enormous width is not onest source for the multiexponential relaxation in $%
Sc $ metal. In order to observe single exponential relaxation it is
necessary to saturate completely nuclear spin system by the comb. But at the
same time the new effect connected with the system of conduction electrons
can appear. Usually, in order to escape the skin-effect manifestations the
NMR experiments in metals are provided with polycrystalline samples and the
crystalline particles are covered by the paraffin for the isolation each
particle from others. If we keep in mind that in the NMR experiments at
liquid helium temperatures such paraffin layer might make difficult the heat
exchange between powder particles and helium bath. So the conduction electron
system being the thermal reservoir for nuclear spin system can have
the temperature difference from the helium bath one just after a comb. The
elementary calculations of the phonon contribution to the molar heat
capacity of $Sc$ metal accordingly to the Debye's model (Debye temperature $%
\Theta _D=352.2K$ \cite{Swenson}) 
\begin{equation}
C_{ph}=\frac{12\pi ^4}5Nk_B\left( \frac T{\Theta _D}\right) ^3  \label{Cph}
\end{equation}
and the electron contribution 
\begin{equation}
C_{el}=\gamma _eT,  \label{Cel}
\end{equation}
$C_{el}=\gamma _eT$, where $\gamma _e=10.38mJ/mol\cdot K^2$ \cite{Swenson},
give us the values listed in Table 3.

One can see that the phonon heat capacity is so small to termalize the
conduction electron system at these temperatures and other thermal reservoir
is necessary, for example, the helium bath. The effects of the taking into
account the finite phonon heat capacity in the nuclear spin-lattice
relaxation problems were considered during many years \cite{Kittel}-\cite
{Eska}. In this section we shall not discuss in the details of the
mechanism of energy transfer from the conduction electrons to the helium
bath (is this path by the phonons in scandium crystal lattice or not, how
the paraffin layer effects on this path and so on). We shall consider only
phenomenologically the heat exchange between conduction electrons and the
helium bath in the following two particular cases:

a) The ''usual'' heat exchange, when the heat flow from conduction electrons
to liquid helium is proportional to the difference between their
temperatures;

b) The heat exchange of the ''Kapitza resistance'' type , when the heat flow
is proportional to the difference of the fourth powers of their temperatures.

In Fig.\ref{fig9} the simplified scheme of the thermal reservoirs and heat flows
taking place in the problem of nuclear magnetic relaxation in metals is
shown. $Q_{1}$ and $Q_{2}$ represent the amounts of heat coming to the
nuclear spin system and the conduction electron system respectively from the
radiofrequency pulses saturating nuclear spin system.

After a comb the heat flows in the considered system can be described by the
following kinetic equations: 
\begin{equation}
\frac{dQ_{n}}{dT}=C_{n}\frac{dT_{n}}{dt}=-\alpha \left( T_{n}-T_{el}\right)
\label{KinN}
\end{equation}
\begin{equation}
\frac{dQ_{el}}{dt}=C_{el}\frac{dT_{el}}{dt}=-\beta \left(
T_{el}-T_{0}\right) -\alpha ^{\prime }(T_{el}-T_{n})\text{ \ \ \ \ \ \ \ \ \
\ \ case a)}  \label{KinElA}
\end{equation}
\begin{equation}
\frac{dQ_{el}}{dt}=C_{el}\frac{dT_{el}}{dt}=-\beta ^{\prime }\left(
T_{el}^{4}-T_{0}^{4}\right) -\alpha ^{\prime }(T_{el}-T_{n})\text{ \ \ \ \ \
\ \ \ \ \ \ case b)}  \label{KinElB}
\end{equation}
Here $\alpha $, $\alpha ^{\prime }$, $\beta $, $\beta ^{\prime }$ are the
kinetic parameters, $T_{n}$, $T_{el}$ and $T_{0}$ are the temperatures of
nuclear spin system, conduction electron system and liquid helium
correspondingly. As far as the heat capacity of the nuclear spin-system is
very small in comparison with the conduction electron one at liquid helium temperature we can neglect the
second term in the equations $\ref{KinElA}$, \ref{KinElB}. The given
equations can be re-written in the more convenient form by introducing the
inverse temperatures of systems - $\beta _{n}$, $\beta _{el}$, $\beta _{0}$
for nuclear spin system, conduction electron system and liquid helium
correspondingly - and the dimensionless quantity $m$ and $n$: 
\begin{equation}
m=1-\beta _{n}/\beta _{0}\text{ \ \ \ \ \ and \ \ \ }n=1-\beta _{el}/\beta
_{0}  \label{MN}
\end{equation}

We should take into account also that the heat capacity of the conduction
electrons depends on temperature linearly. The final system of kinetic
equations looks as follows: 
\begin{eqnarray}
\frac{dm}{dt} &=&\frac 1{\tau _{nel}}\frac{n-m}{1-n}  \label{KinA} \\
&&\text{ \ \ \ \ \ \ \ \ \ \ \ \ \ \ \ \ \ \ \ \ \ \ \ \ \ \ \ \ \ \ \ \ \ \
\ \ \ \ \ \ \ \ \ \ \ \ \ \ \ \ \ \ case a)}  \nonumber \\
\frac{dn}{dt} &=&-\frac n{\tau _{el0}}  \nonumber
\end{eqnarray}
\begin{eqnarray}
\frac{dm}{dt} &=&\frac 1{\tau _{nel}}\frac{n-m}{1-n}  \label{KinB} \\
&&\text{ \ \ \ \ \ \ \ \ \ \ \ \ \ \ \ \ \ \ \ \ \ \ \ \ \ \ \ \ \ \ \ \ \ \
\ \ \ \ \ \ \ \ \ \ \ \ \ \ \ \ \ \ case b)}  \nonumber \\
\frac{dn}{dt} &=&-\frac 1{\tau _{el0}^{\prime }}\frac{1-(1-n)^4}{1-n^2} 
\nonumber
\end{eqnarray}

In order to obtain the temporal dependence of the nuclear magnetization it
is necessary to find the time dependence of the quantity $m(t)$ from \ref
{KinA} or \ref{KinB} and then take into account that nuclear magnetization $%
M$ is proportional to the inverse nuclear spin temperature $\beta _n$ (Curie
law). The non-linear systems of kinetic equations \ref{KinA} and \ref{KinB} can be
solved numerically for the different values of the kinetic parameters $\tau
_{nel}$, $\tau _{el0}$ and $\tau _{el0}^{\prime }$ and initial deviations of
the conduction electron temperature from the equilibrium meaning (the
initial value of the parameter $n$). The results of some numerical solutions for the
time evolution of the quantity $m(t)$ are given in Fig.\ref{fig10} and Fig.\ref{fig11}.
As the result of calculation we see that due to the heating of the
conduction electron system by radiofrequency pulses the nuclear magnetic
relaxation has essentially non-single exponential character.

\section{Conclusion.}
The nuclear magnetic relaxation in scandium metal was investigated at liquid
helium temeprature for both single crystal and powder samples. The obtained temperature
and field dependence of the Korringa constant can be described by the means
of the spin fluctuation theory. It was shown the most probable reason for the
high field decreasing of the relaxation rate is the quenching of spin fluctuations.
The other possible reason - the magnetic impurity influence - was also considered.
The reasons of the existing variations of the Korringa constant were discussed. The possible sources
of non-single exponential relaxation were investigated.

\begin{table}
\caption{The values of the Korringa constant obtained from NMR experiments
at different conditions}
\label{table1}
\begin{tabular}{|c|c|c|c|c|c|}
Sample, & Temperature, & Frequency, & Magnetic
& $T_1T$, & Ref. \\
purity& $K$ & $MHz$ & field, $T$ & $\sec \cdot K$ & \\  
angle between &  &  &  &  &  \\
magnetic field &  &  &  &  &  \\
and $c$-axis &  &  &  &  &  \\  
\tableline polycrystalline, & $1.7\div 300$ & $14.5$ & $1.4$ & $0.6$ & \cite{Blumberg}
\\ 
$99.9\%$ & $1.5\div 77.0$ & $10.0$ & $0.967$ & $0.11$
& \cite{Masuda1} \\ 
\tableline single crystal, & $77.0$ & $8.0$ & $0.77$ & $%
(1.25-0.58\sin ^2\theta )^{-1}$ & \cite{Fradyn} \\
$99.9\%$ &  &  &  &  &  \\
$\theta $ &  &  &  &  &  \\  
\tableline polycrystalline & $1.0\div 4.0$ & $17.0$, $30.0$ & $1.64,$ 
$2.89$ & $1.3\pm 0.2$ & \cite{Narath}, \cite{NarathLa} \\ 
\tableline polycrystalline, & $1.5\div 77.0$ & $10.0$ & $%
0.967$ & $1.5\pm 0.2$ & \cite{Masuda2} \\
$99.98\%$ &  &  &  &  &  \\ 
\tableline single crystal, $\cos \theta =1$ & $77$ & $8.0$ & $0.77$ & $0.81$
&  \\ 
single crystal, $\cos \theta =0$ & $77$ & $8.0$ & $0.77$ & $1.49$ & \cite
{Ross} \\ 
single crystal, $\cos \theta =1/\sqrt{3}$ & $4.2$ & $8.0$ & $0.77$ & $1.1\pm
0.1$ &  \\ 
polycrystalline & $300$ & $12.0$ & $1.16$ & $1.2\pm 0.1$ &  \\ 
\tableline & $1.3\div 300$ & $50.0$ & $4.82$ & $1.6$ &  \\ 
& $1.3\div 4.2$ & $20.0$ & $1.93$ & $1.48$ & \cite{Masuda3} \\ 
& $1.3\div 4.2$ & $5.0$ & $0.48$ & $1.36$ &  \\ 
\tableline & $0.003\div 0.01$ & $0.39$, $0.26$ & $0$ (NQR) & $0.09\pm 0.009$
& \cite{Cornell} \\  
\end{tabular}
\end{table}

\begin{table}
\caption{The fitting parameters for the measurements of $M/H$}
\label{table2}
\begin{tabular}{||c|ccc|ccc||}
Orientation of the&  & Eq. \ref{exp} &  &  & Eq. 
\ref{quad} &  \\
external magnetic field &  &  &  &  &  &  \\  
\tableline & $a_1$, $emu/g$ & $b_1$, $emu/g$ & $c_1$, $T$ & $a_2$, $emu/g$ & 
$b_2$, $emu/g$ & $c_2$, $T^{-2}$ \\ 
\tableline along $a$-axis & $8.18\cdot 10^{-6}$ & $1.14\cdot 10^{-6}$ & $0.99
$ & $8.17\cdot 10^{-6}$ & $1.04\cdot 10^{-6}$ & $1.70$ \\ 
\tableline along $c$-axis & $7.75$ & $1.47$ & $1.24$ & $7.75$ & $1.33$ & $%
1.05$ \\ 
\end{tabular}
\end{table}

\begin{table}
\caption{The phonon and electronic heat capacities at low temperatures in $%
^{45}Sc$}
\label{table3}
\begin{tabular}[t]{||c|c|c||}
$T,K$ & $C_{ph},10^{-3}J/(K\cdot mol)$ & $C_{el},10^{-3}J/(K\cdot mol)$ \\ 
\tableline $4.2$ & $3.1$ & $44.9$ \\ 
\tableline $1.5$ & $0.14$ & $16.1$ \\ 
\end{tabular}
\end{table}

\begin{figure}
\caption{Cw $^{45}Sc$ NMR spectrum ($\nu =5.113MHz$)in polycrystalline sample at liquid helium temperature.
\label{fig1}}
\end{figure}

\begin{figure}
\caption{Free induction decay of $^{45}Sc$ signals at $T=1.5K$. The frequencies and pulse durations are shown.
\label{fig2}}
\end{figure}

\begin{figure}
\caption{The transverse magnetization decays for $Sc$ single crystal and polycrystalline samples.
\label{fig3}}
\end{figure}

\begin{figure}
\caption{The dependencies of $^{45}Sc$ nuclear longitudinal relaxation time on the pulse duration time at $T=1.5K$.
The resonance frequencies are shown.
\label{fig4}}
\end{figure}

\begin{figure}
\caption{The recovery of the longitudinal magnetization in pulse $^{45}Sc$ NMR experiments with single crystal
at $T=1.5K$. The pusle duration time $\tau =1\mu s$.
\label{fig5}}
\end{figure}

\begin{figure}
\caption{The magnetic field
dependence of the Korringa constant. Tha data known from the literature as well as our data
are shown. The solid line represents the result of the fitting procedure by use Eq.\ref{SCR}.
The dashed line is drawn with taking into account the magnetic impurity influence
(see text). 
\label{fig6}}
\end{figure}

\begin{figure}
\caption{The magnetic field dependence of the parameter $\zeta $. The solid and dashed lines
show the fitting by use Eq. \ref{quad} and Eq. \ref{exp} correspondingly.
\label{fig7}}
\end{figure}

\begin{figure}
\caption{The time evolution of nuclear magnetization at different initial saturation.
The nuclear spin-lattice ralaxation time is shown.
\label{fig8}}
\end{figure}

\begin{figure}
\caption{The thermal reservoirs and heat flows in the system under consideration.
\label{fig9}}
\end{figure}

\begin{figure}
\caption{The time evolution of $m(t)$. The open symbols correspond to the case of
the "usual" heat exchange, whereas the solid ones correspond to the "Kapitza resistance"
type of the heat exchange. The kinetic parameters are the following: $\tau _{nel}=1sec$,
$\tau _{el0}=\tau _{el0}^{\prime }=2sec$. The initial conditions for the electron system
are shown in the frame. The nuclear spin system is completely saturated at initial time. 
\label{fig10}}
\end{figure}

\begin{figure}
\caption{The same as in Fig.10. $\tau _{el0}=\tau _{el0}^{\prime }=10sec$. 
\label{fig11}}
\end{figure}

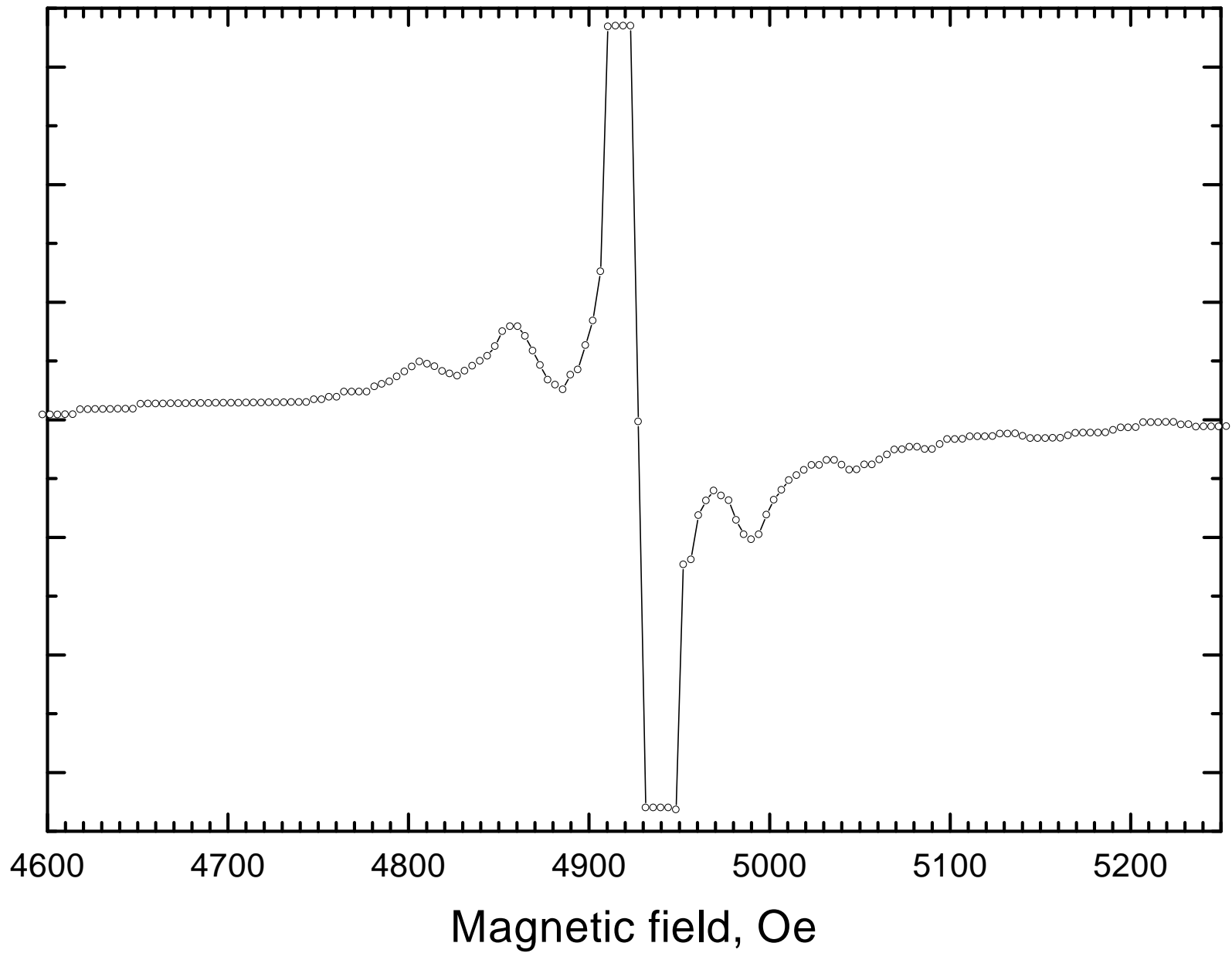
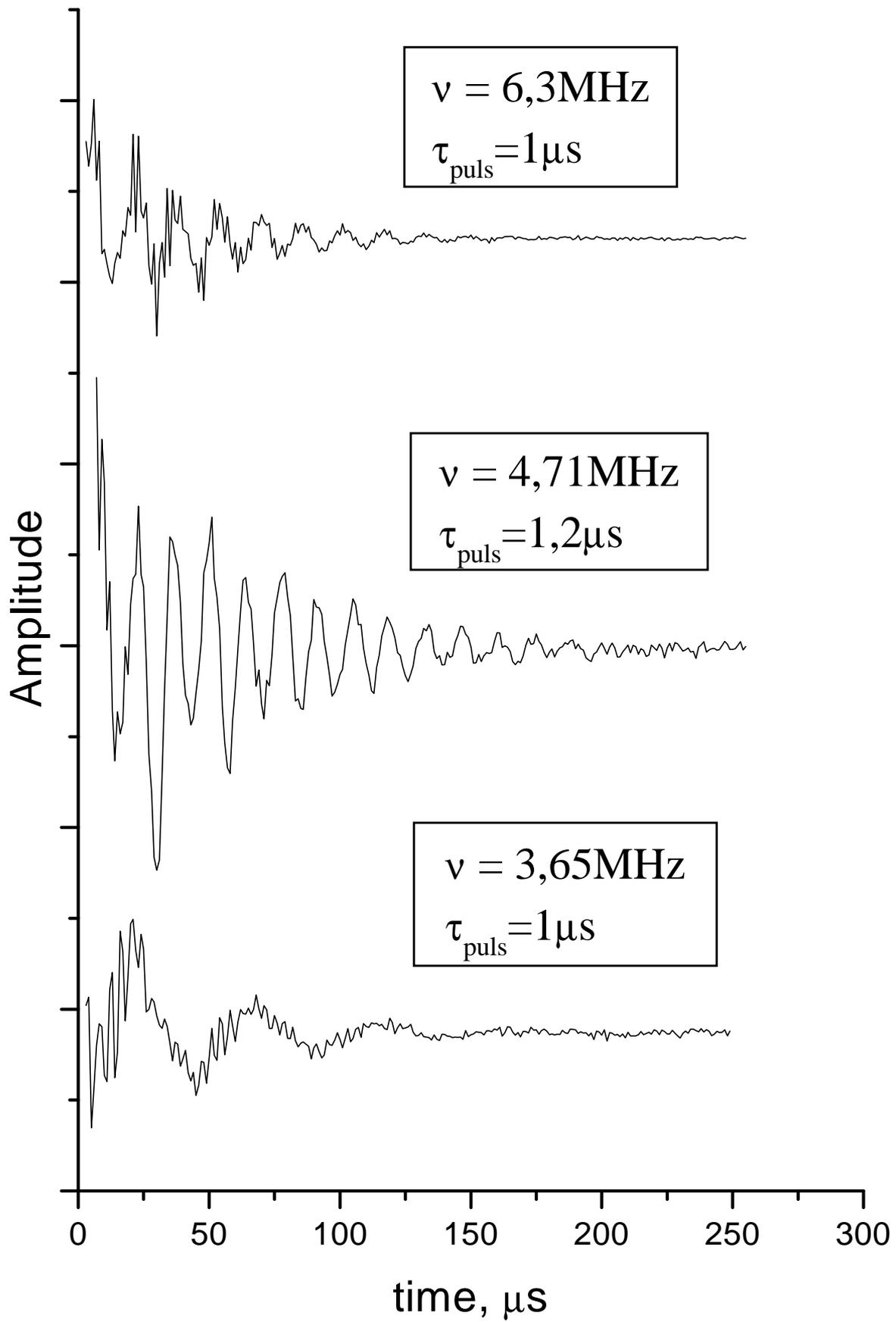
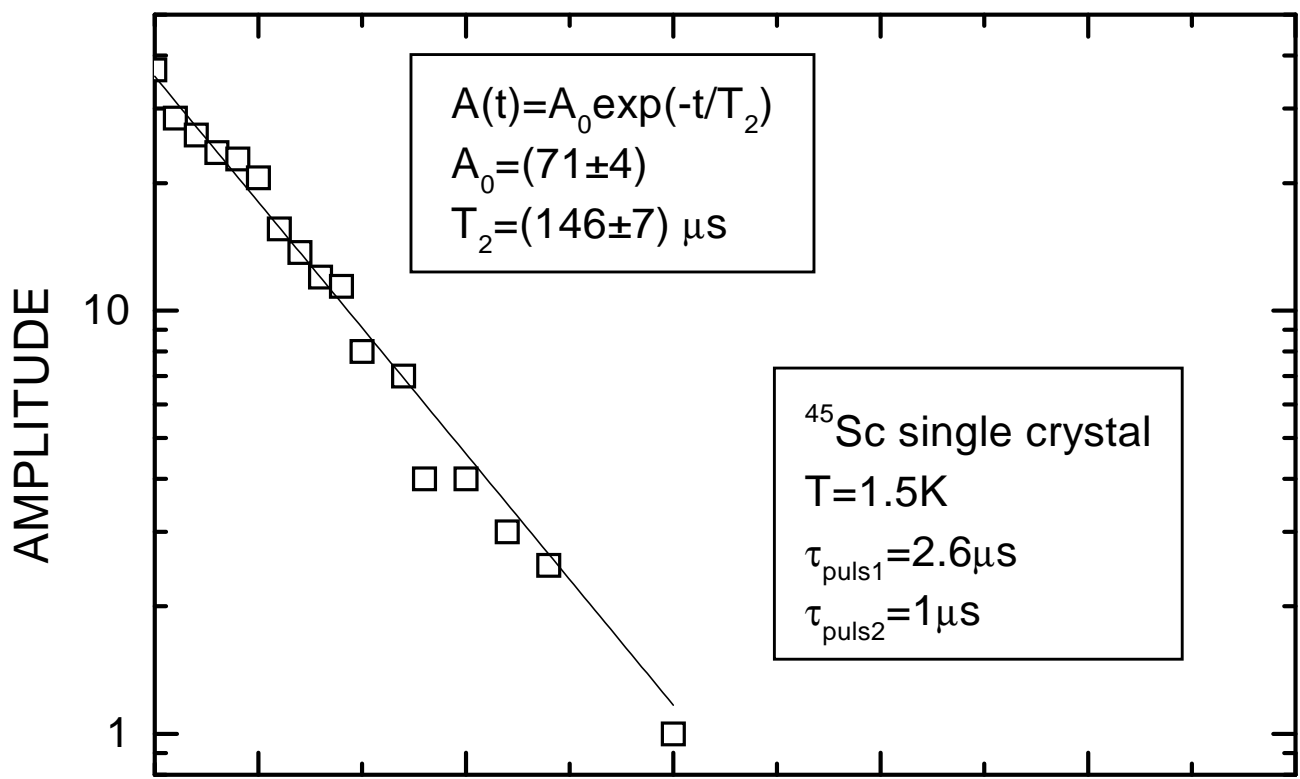
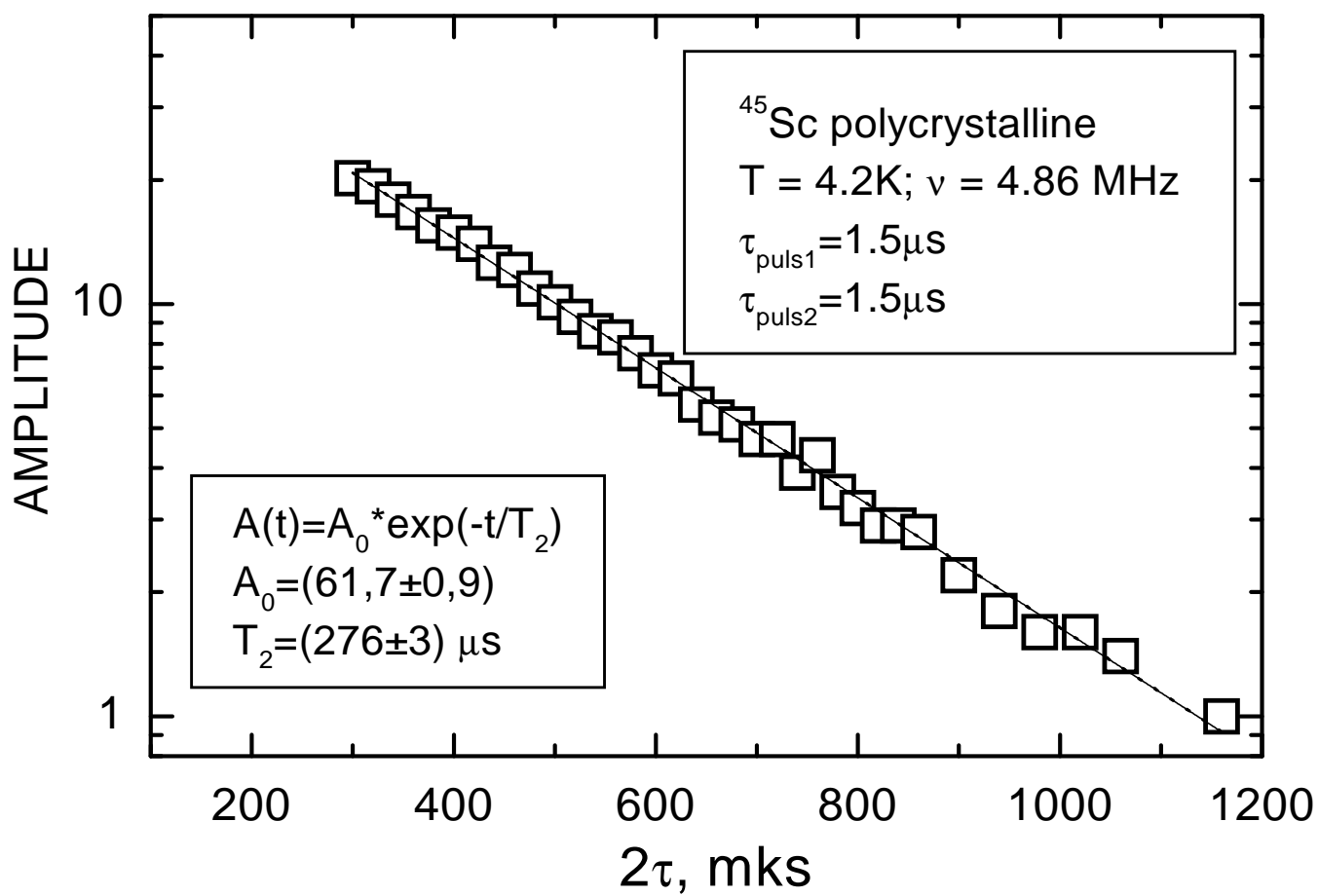
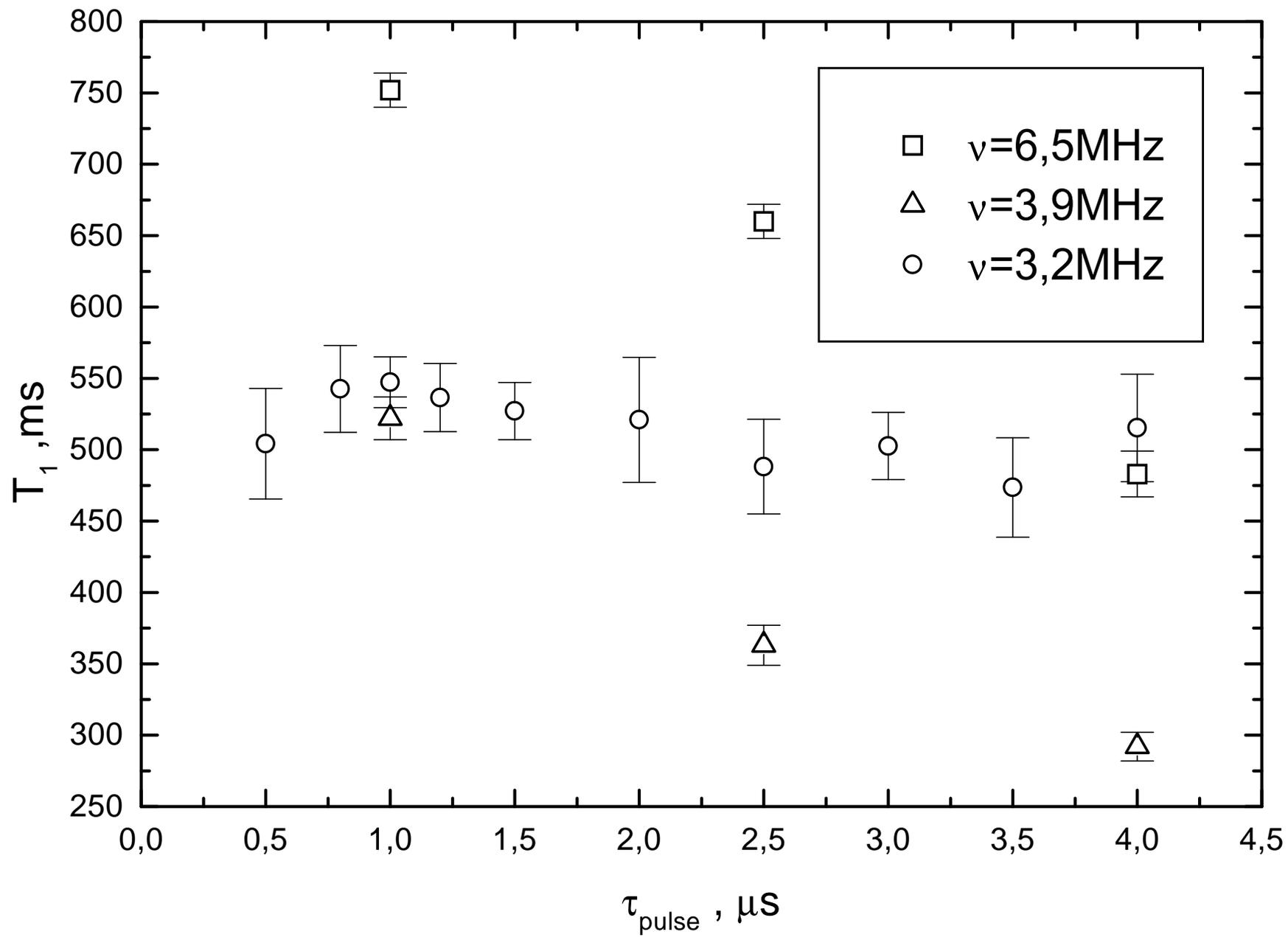
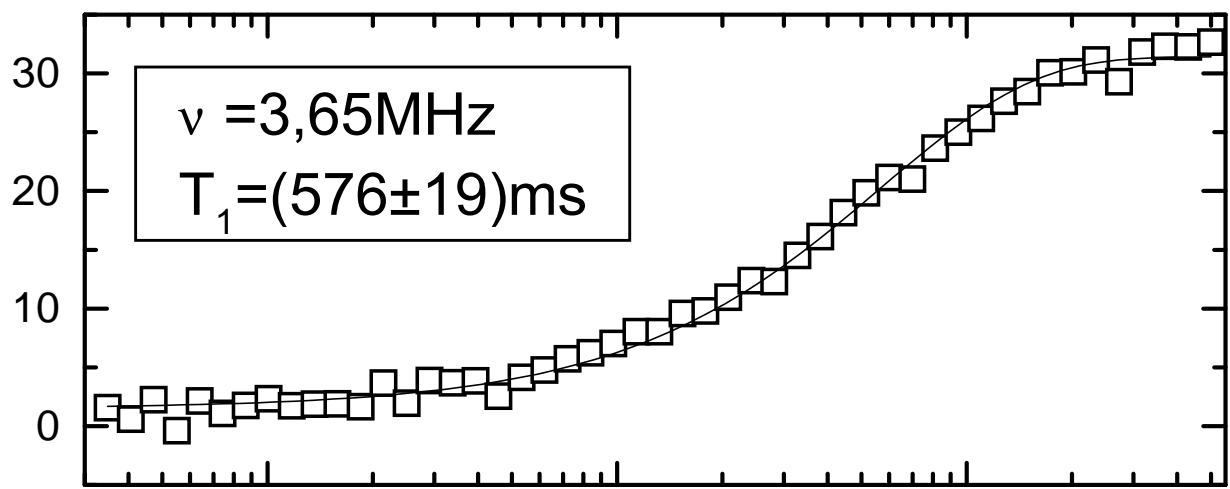
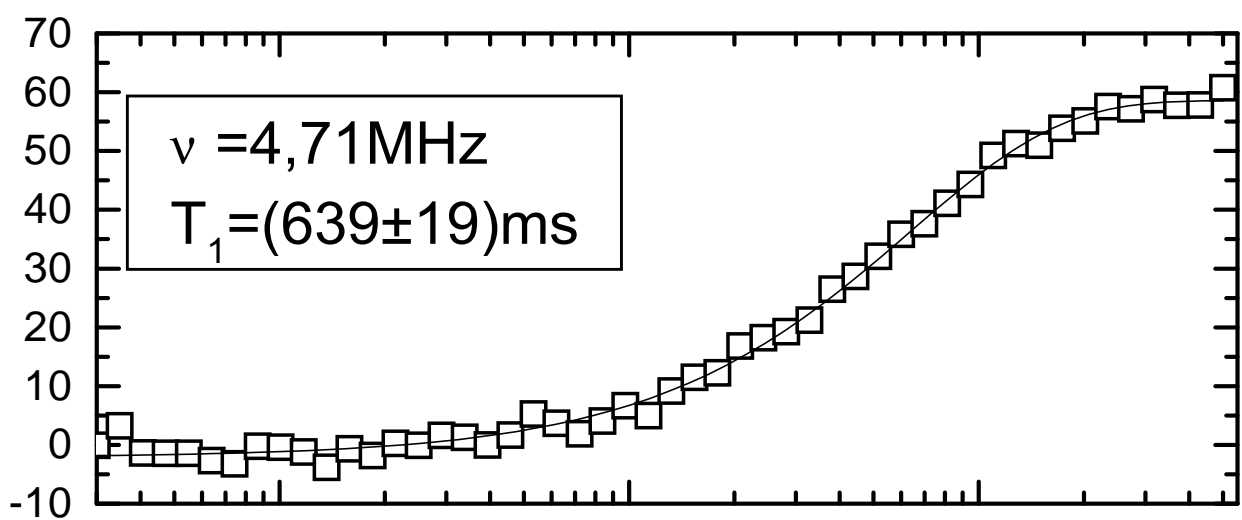
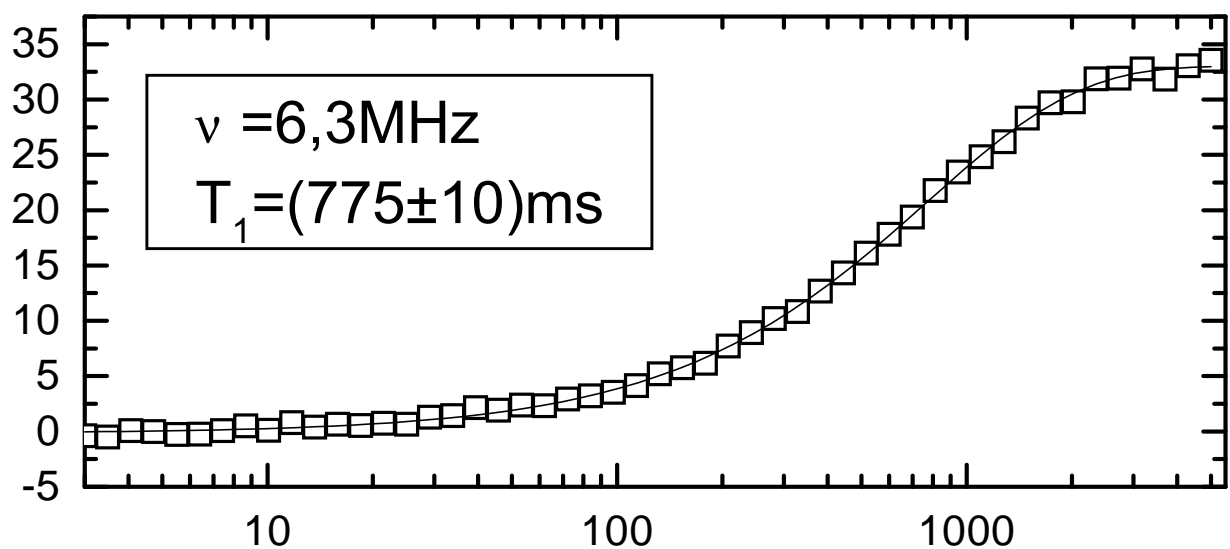
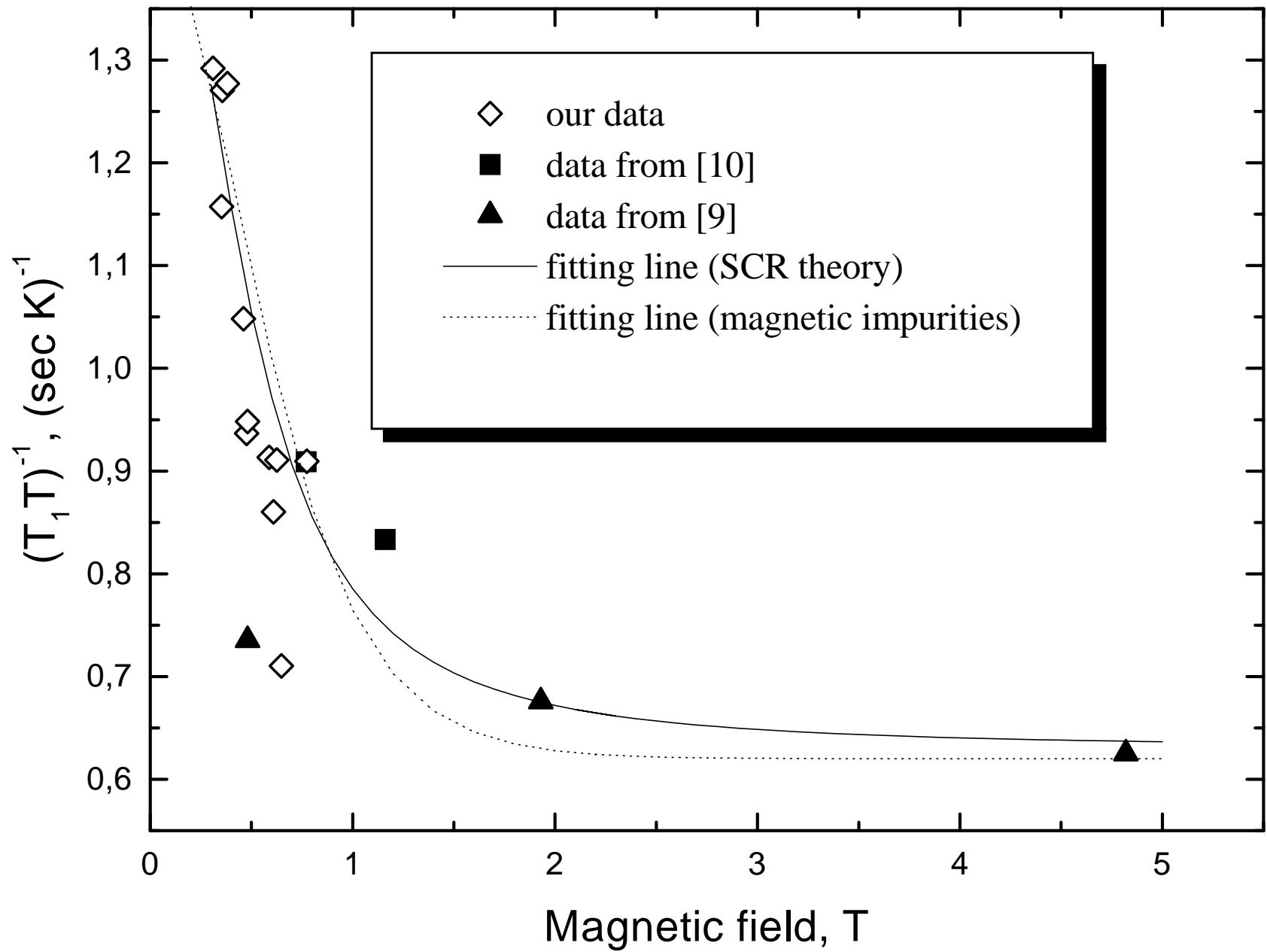
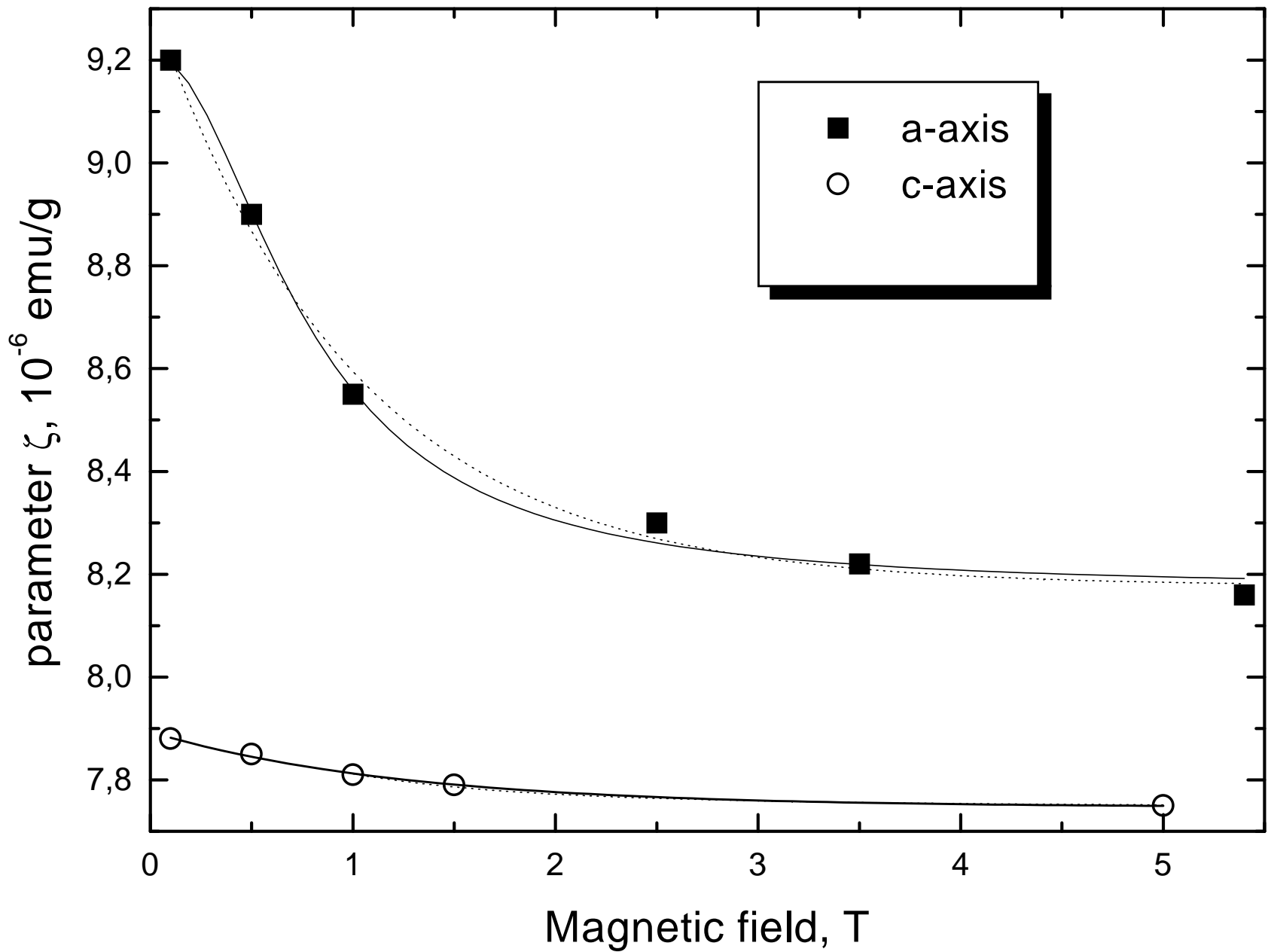
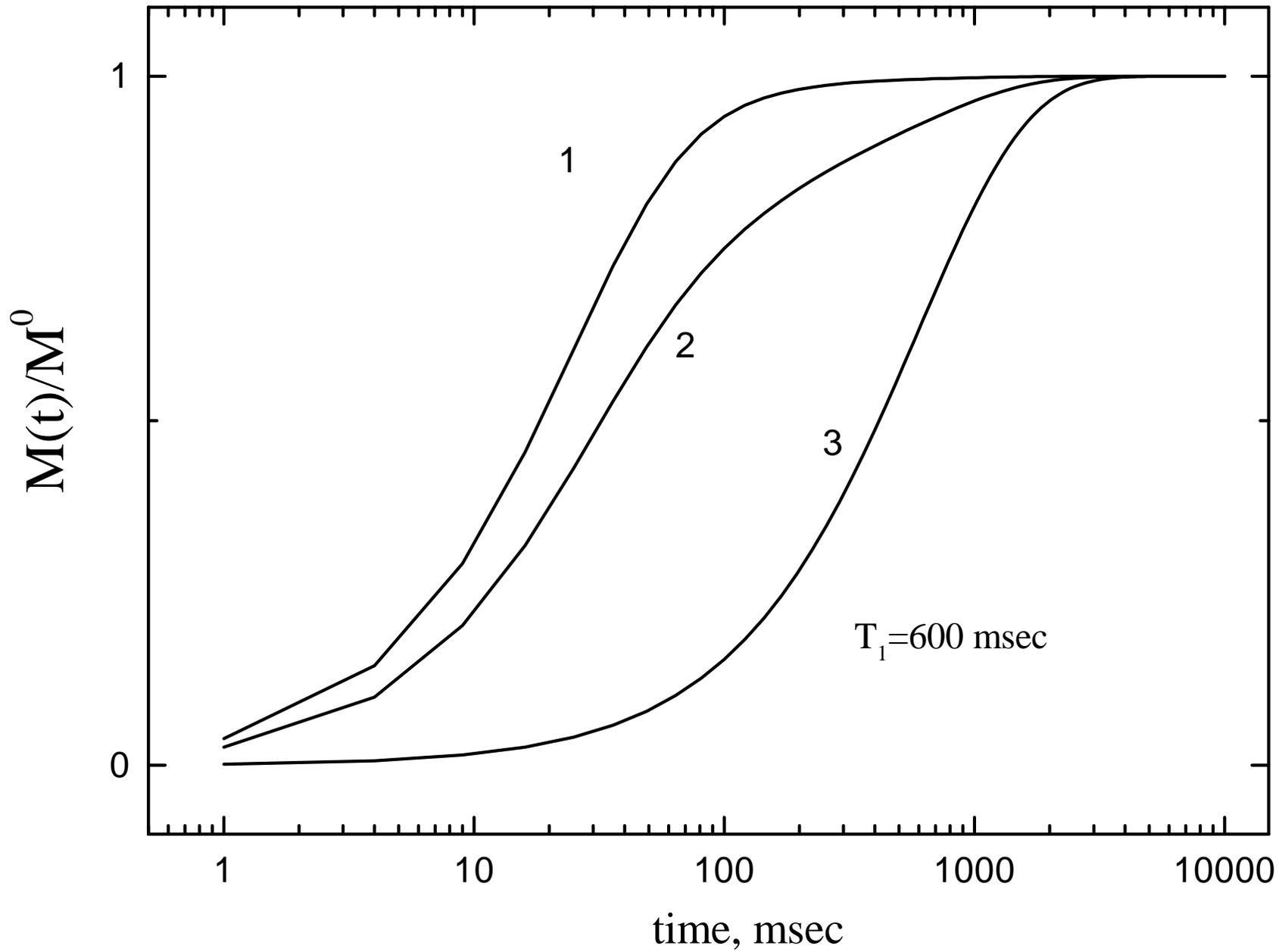
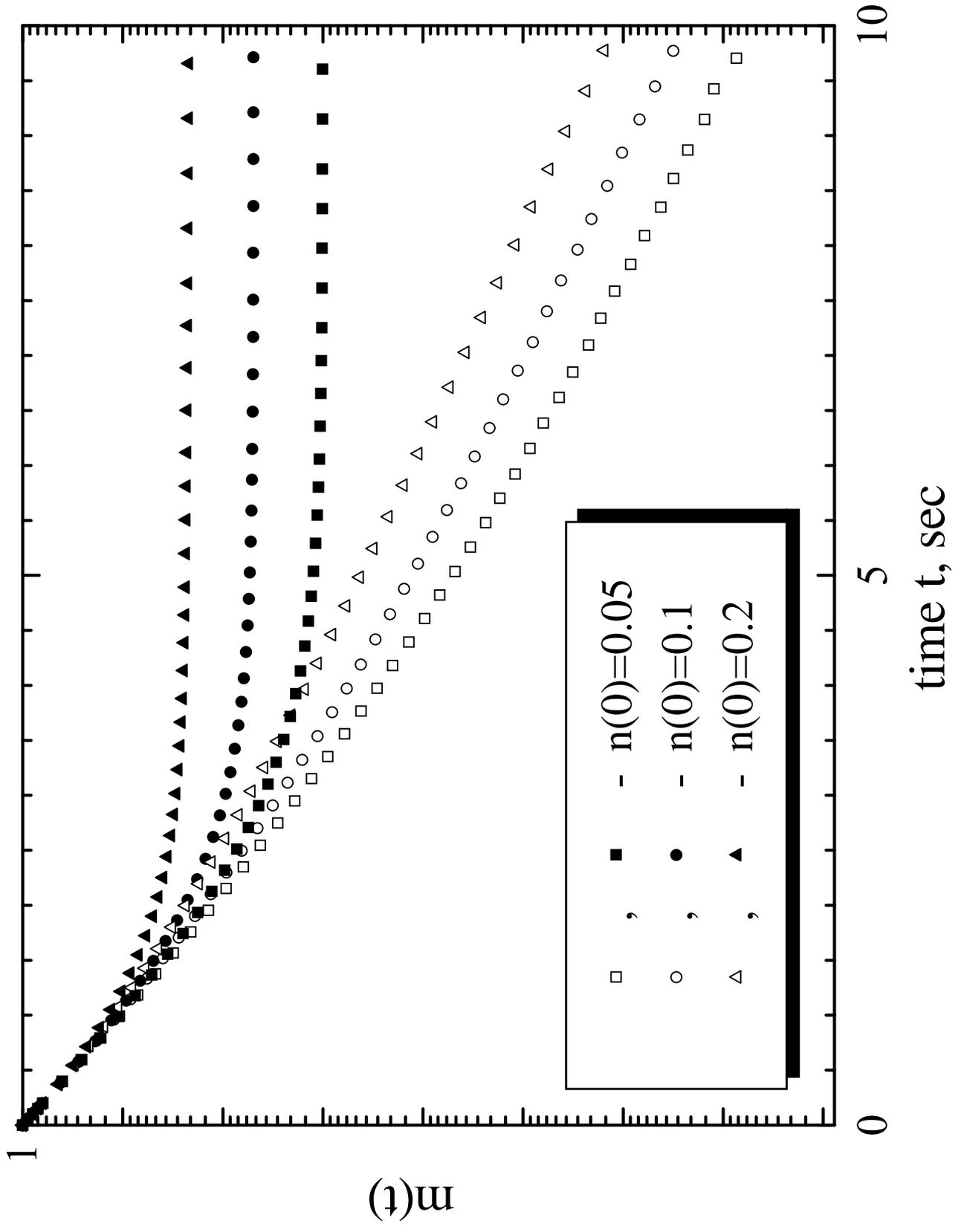
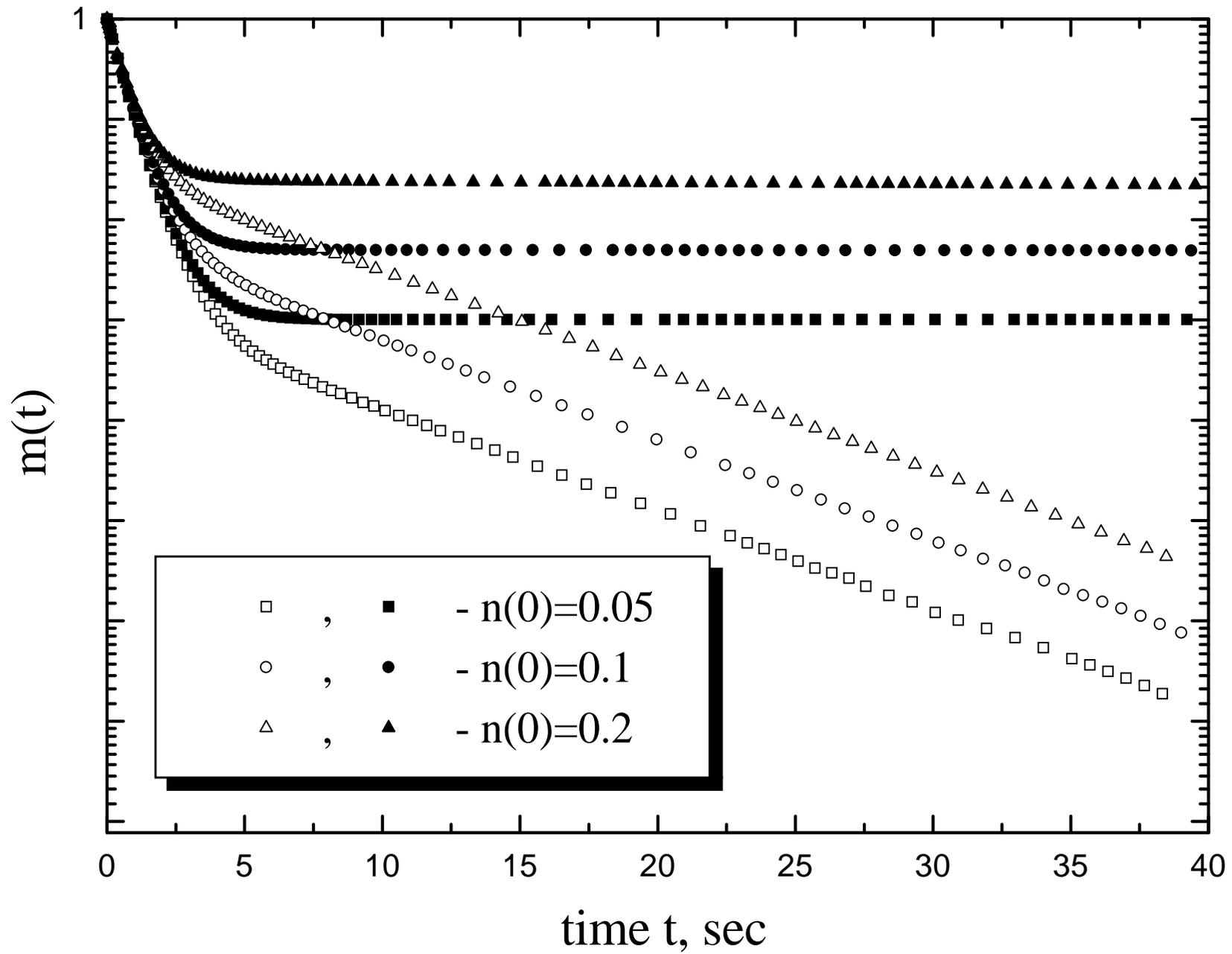
\end{document}